\documentclass[nofootinbib,twocolumn,preprintnumbers,amsmath,amssymb]{revtex4}

\begin{document}
\title{Solar system tests \textit{do not} rule out 1/R gravity}

\author{Qasem \surname{Exirifard}}
\affiliation{
School of Physics, Institute for Research in Fundamental Sciences (IPM), P.O.Box 19395-5531, Tehran, Iran}

\email{exir@theory.ipm.ac.ir}

\begin{abstract}
We argue that Solar system tests do not rule out 1/R gravity at least due to the reason addressed in Phys. Rev. D 74 (2006) 121501 [astro-ph/0610483] (ref. [1]) and subsequent published papers. 
Ref. [1] has not only modified the Einstein-Hilbert action but also has changed the boundary conditions since they altered the equations of motion. In Einstein-Hilbert action equations are second order, so the fall off of the fields suffices to single out a unique solution. In 1/R gravity the equations are fourth order, so we should impose additional boundary conditions. Perhaps the boundary condition we must impose is that the abrupt change in the second derivative of the metric near the surface of the Sun remains intact by adding `1/R' corrections to the equations of motion. The solution of 1/R gravity with this boundary condition remains consistent with the solar system tests. 
Ref. [1] assumes that as soon as they perturbatively modified the equations then the Ricci scalar becomes smooth on the surface of the Sun. This assumption is simply wrong because the boundary conditions and equations of motions are two different entities.

\end{abstract}

\maketitle
There exists a debate in the literature on the consistency of 
\begin{equation}\label{R-action}
S= \frac{1}{16 \pi G} \int d^4 x \sqrt{- \det g} (R - \frac{\mu^4}{R}) + \int d^4x \sqrt{-\det g } L_{m}
\end{equation}
for $\mu^{-1}\approx 10^{26} meters$ with the Solar system tests, for example look at \cite{Erickcek:2006vf,Chiba:2003ir,Kainulainen:2007bt,Olmo:2006eh,DeDeo:2007yn,Rajaraman:2003st,Multamaki:2006zb,Faraoni:2006hx,Ruggiero:2006qv,Allemandi:2005tg}. In particular ref. \cite{Erickcek:2006vf} claims  that properly matching the metric inside and outside the Sun rules out $\frac{1}{R}$ gravity. A sharp inspection of  \cite{Erickcek:2006vf}, however, reveals that \cite{Erickcek:2006vf} has not properly studied the equations. This was also noticed by \cite{DeDeo:2007yn}. In the following we would like to clarify why the conclusion of \cite{Erickcek:2006vf} is not right in addition to demonstrating the source of this wrong conclusion.\footnote{Common accepted misunderstandings about $f(R)$ among referees provide a ground for rejecting my works. In order to clear these misunderstandings, I have decided to publish this short note.}

Contracting the equations of motion of \eqref{R-action} with the inverse of the metric yields
\begin{equation}\label{R-eq}
\Box \frac{\mu^4}{R^2} -\frac{R}{3} + \frac{\mu^4}{R} = \frac{8 \pi G T}{3}\,,
\end{equation}
where $T= g^{\mu\nu} T_{\mu\nu}$ and the speed of light is set one. Now let us define a new variable, $x$, through $R=- 8 \pi G x$. Rewriting \eqref{R-eq} in terms of $x$ and rearranging the terms yields
\begin{equation}\label{x-eq}
x = 1 - \frac{3 \mu^4}{(8 \pi G T)^2} (- \frac{1}{x} + \frac{1}{8 \pi G T} \Box \frac{1}{x^2} + O(\nabla T))\,.
\end{equation}
Note that the Einstein-Hilbert gravity holds $x=1$. We can obtain the order of magnitude of $T$ for the Sun by $T \approx \frac{3 M_\odot}{4 \pi R_\odot^3}$ where $R_\odot$ is the radius of the Sun. Knowing the order of magnitude of $T$, we can obtain the order of magnitude for deviation from $x=1$ in \eqref{x-eq}:
\begin{equation}
 \frac{3 \mu^4}{(8 \pi G T)^2} \approx  \frac{3 \mu^4}{(\frac{3 G M_\odot}{R_\odot^3 (\text{speed of light})^2})^2}= \frac{4 \mu^4 R_\odot^6}{3 r_\odot^2}\,,
\end{equation}
where $r_\odot = \frac{2 G M_\odot}{(\text{speed of light})^2}$ is the Schwarszchild radius associated to the mass of the Sun. Using $R_\odot = 1.39 \times 10^ 9 meters$, $r_\odot = 3 km$, the order of magnitude of the deviation from $x=1$ in \eqref{x-eq} reads
\begin{equation}
 \frac{3 \mu^4}{(8 \pi G T)^2} \approx  10^{-55}.
\end{equation}
Note that the order of magnitude of the coefficient in the front of $\Box \frac{1}{x^2}$ and similar terms in \eqref{x-eq} is $\frac{R_\odot^3}{r_\odot} \times \frac{3 \mu^4}{(8 \pi G T)^2}= 10^{-31} meters^2$. Therefore even the non-homogeneousity of the matter's distribution in the Sun does not produce a significant deviation from $x=1$ inside the Sun. 
Recalling that the Einstein-Hilbert gravity holds $x=1$ besides  extraordinarily  small deviation from $x=1$, we conclude that  the Einstein-Hilbert action quite-perfectly describes the physics in and outside the Sun for \eqref{R-action}.

Now let us inspect what leads  \cite{Erickcek:2006vf} to the wrong conclusion. Ref. \cite{Erickcek:2006vf} has defined a function by
\begin{equation}
c = -\frac{1}{3} + \frac{\mu^4}{R^2}\,,
\end{equation}
which we refer to as the C-function. The authors  then have taken it granted that the C-function encodes the deviation from the vacuum solution even outside the matter's distribution. This means that the authors have failed to realize that  the C-function can be identically zero outside the matter's distribution.  In other words the deviation from the vacuum solution might  be encoded in other scalars rather the  Ricci scalar or equivalently the C-function.  They than have rewritten \eqref{R-eq} in terms of the C-function
\begin{equation}\label{c-eq}
\Box c + \frac{\mu^2 c}{\sqrt{c+ \frac{1}{3}}} = \frac{8 \pi G  T}{3}, 
\end{equation}
before approximating it to
\begin{equation}\label{electro-eq}
\nabla^2 c = \frac{8 \pi G T}{3}\,.
\end{equation}
It then appears that the authors have assumed that the C-function and its derivatives are continuous on the surface of the Sun. But in the Einstein-Hilbert gravity  what remains continuous on the boundaries are the metric and its first derivatives. For example we know that the Ricci scalar or equivalently the C-function is not continuous on the boundary. Therefore instead of choosing C-function as what \cite{Erickcek:2006vf} has chosen, we must choose it in the following way
\begin{itemize}
\item Outside the Sun ($r>R_\odot$), $c=0$. Note that $c=0$ solves the equation outside the Sun.
\item Inside the Sun we must find a solution of \eqref{c-eq} that remains bounded inside the star. While the metric and its first derivatives are continuous on the surface of the Sun.
\end{itemize} 
The above choice means that within the electrostatic approximation to the equations - \eqref{electro-eq}-, the surface of the Sun effectively plays the role of a conducting surface accommodating some amount of  `charge' that completely cloaks the `charge' inside the Sun. Setting $C=0$ outside the Sun leads to the  Schwarszchild metric in the Solar system which is in agreement with observation. Therefore 1/R gravity is not ruled out at least due to reason addressed in  \cite{Erickcek:2006vf}.

\end{document}